\documentclass{ws-p8-50x6-00}

\begin{document}

\title{Observational signatures of the first stars}

\author{Aparna Venkatesan}

\address{CASA, Department of Astrophysical and Planetary Sciences, \\
University of Colorado, 389 UCB, Boulder, CO 80309-0389, USA \\
E-mail: aparna@casa.colorado.edu}

\author{James W. Truran}

\address{Department of Astronomy and Astrophysics, 5640 S. Ellis Ave., \\
University of Chicago, Chicago, IL 60637, USA \\ E-mail:
truran@nova.uchicago.edu}

\maketitle

\abstracts{At present, there are several feasible observational probes
of the first stars in the universe. Here, we examine the constraints
on early stellar activity from the metallicity of the high-redshift
Ly-$\alpha$ clouds, from the effects of stellar ionizing photons on
reionization and the cosmic microwave background (CMB), and from the
implications of gravitational microlensing results for the presence of
stellar remnants in galactic halos. We also discuss whether the above
signatures are consistent with each other, i.e., if they reflect the
same population of stars.}

The epoch and sites of formation of the first stars in the universe
are some of cosmology's most intriguing questions today. Observations
of the most distant quasars and galaxies to date imply that hydrogen
reionization of the intergalactic medium (IGM) occurred before
redshifts, $z$ $\sim$ 6. While a variety of astrophysical objects can
reionize the IGM, the apparent decrease in the space density of
quasars beyond $z \sim 3$ implies that early stars must play some role
in reionization; furthermore, they may account for the persisting
trace levels of metal enrichment seen in the high-$z$ Ly-$\alpha$
clouds (see Venkatesan\cite{v00} for the relevant references). Here,
we will focus on three specific probes of early stellar activity:
stellar remnants, which may contribute to galactic halo dark matter
(DM) in the form of Massive Compact Halo Objects (MACHOs); the
influence of stellar ionizing photons on the CMB; and, the metal
enrichment of the IGM. In particular, one may ask how each of these
constrains $f_\star$, the fraction of baryons that participate in any
early episodes of star formation. \\ \\ $\bullet$ {\bf Stellar
Remnants}: Recent observations of gravitational microlensing towards
the Large and Small Magellanic clouds from several collaborations,
including MACHO and EROS, indicate that the lenses, if they are
present in the Galactic halo, have a most likely mass $\sim$ 0.5--1
M$_\odot$, and a halo mass fraction of about 0.2--0.3 (see, e.g.,
Kerins\cite{kerins}, and references therein). This has renewed
theoretical interest in the viability of baryonic DM candidates, which
include white dwarfs (WDs; see, e.g., Fields, Mathews \&
Schramm\cite{fields97}), and neutron stars/black holes (NSs/BHs;
Venkatesan, Olinto \& Truran\cite{v99}; see also Fields, Freese \&
Graff \cite{fields98} for a review of various models for baryonic
DM). Although these models (just barely) satisfy a number of
constraints, almost all of the baryons must either participate in an
early burst of star formation that creates the MACHOs, or assist in
dilution of the stellar ejecta to avoid overpollution in $^4$He or
metals; in effect, $f_\star$ $\geq$ 0.8, when dilution requirements
are included.  \\ \\ $\bullet$ {\bf Stellar Radiation and the CMB}:
The first stars generate ionizing photons that increase the free
electron population in the IGM; these electrons then interact with the
CMB via Thomson scattering, which results in an overall damping of the
CMB's primary temperature anisotropies. Also created are new secondary
temperature anisotropies and a linear polarization signal, which is a
relatively clean probe of the reionization epoch. Analysis of CMB data
can yield constraints on a chosen set of cosmological parameters (CPs)
and the optical depth to reionization, $\tau_e$. If we further adopt a
stellar reionization model for $\tau_e$, then well-known degeneracies
in CMB parameter extraction can be broken; also, limits on CPs can be
used to constrain astrophysical quantities such as $f_\star$ (see
Venkatesan\cite{v00}, and references therein). Strong constraints on
$f_\star$ within individual star-forming halos are possible, in
principle, by this method with polarization data from future CMB
experiments such as $PLANCK$ (see, e.g., Figure 13 of
Venkatesan\cite{v00}).  \\ \\ $\bullet$ {\bf Metal Production}: The
IGM at $z \sim$ 3 has an average enrichment, detected in carbon, of
$Z_{\rm IGM}$ $\sim 10^{-2} Z_\odot$; this requires $f_\star$ $\sim$
1\%, if we assume that most of the baryons at $z \sim 3$ are in the
Ly-$\alpha$ clouds. It can be shown that this value of $Z_{\rm IGM}$
implies an average number of stellar ionizing photons per baryon in
the universe, $N_{\rm Lyc}/b$ $\sim$ 10, which again points to a
possibly significant part played by the first stars in
reionization. Note, however, that if this calculation were repeated
using only the carbon (and not total metal) yields from low-$Z$
massive stars, then $N_{\rm Lyc}/b$ $\sim$ 360!  This implies that
either the reionizing stars made a negligible contribution to the IGM
carbon detected at $z \sim 3$, or that in enriching the IGM, they
generated a tremendous amount of ionizing radiation. \\ There are
other drawbacks to relating stellar ionizing activity to $Z_{\rm IGM}$
through carbon: the photons relevant for reionization come from
massive stars ($\geq$ 10 M$_\odot$), while carbon is produced
dominantly by 2--6 M$_\odot$ stars. Thus, different regions of the
stellar IMF are being probed, and may not be mutually constraining if
the IMF was different in the past. Secondly, for burst-driven star
formation (i.e., not continuous), most of the carbon is produced an
order of magnitude in time {\it after} the epoch, set by the
timescales associated with massive stars, of ionizing photons and Type
II supernovae (which could drive galactic outflows). The question then
arises of the mechanism by which the carbon is expelled from
individual halos and distributed so ubiquitously in the IGM by $z \sim
3$.

{\bf To summarize}: [1] In order to explain the observations of MACHOs
as stellar remnants, most of the baryons must either participate in an
early burst of star formation or assist in the dilution of enriched
stellar ejecta to $\sim$ solar levels. For both WD and NS/BH MACHO
models, the required $f_\star$ far exceeds that needed for stellar
reionization. [2] Strong limits on $f_\star$ from reionization alone
are possible with polarization data from future CMB experiments such
as $PLANCK$, but this may be hampered by foregrounds. [3] Combining
reionization and metal enrichment constraints, it is possible for the
same population of stars to reionize the universe in hydrogen by $z
\sim$ 6, and produce $Z_{\rm IGM}$ $\sim 10^{-2} Z_\odot$ in carbon at
$z \sim 3$, if $f_\star$ $\sim$ 1--2\% and star formation is
continuous.  [4] If star formation occurred in bursts, it is unclear
how to relate ionizing stellar activity (from short-lived massive
stars) to the carbon abundance in the IGM (from intermediate-mass
stars). The detection of elements which are the products of massive
stars, e.g., oxygen or silicon, in absorption-line systems at $z \geq$
3, and a comparison of the values of $Z_{\rm IGM}$ implied by their
abundances with that measured in carbon should help to consistently
address whether the reionizing stellar population generated the metals
observed in the high-$z$ IGM. The details of the calculations on this
connection can be found in our forthcoming work (Venkatesan \&
Truran\cite{v01}).

\section*{Acknowledgments}
A.V. gratefully acknowledges support from NASA LTSA grant NAG 5-7262.

\end{document}